\documentstyle [12pt]{article}
\begin{document}
\baselineskip 0.75cm
\begin{center}
\ \\
\vspace{0.7cm}
{\large{\bf Universality of Symmetry and Mixed-symmetry
Collective Nuclear States}}
\  \\
\vspace{0.5cm}
\  \\
{Bradley H. Smith$^{1,2}$, Xing-Wang Pan$^{1}$, Da Hsuan Feng$^{1}$
and Mike Guidry$^{2,3}$}\\
\vspace{0.4cm}
{\small{\em $^{1}$Department of Physics, Drexel University,
Philadelphia, PA 19104--9984}} \\
{\small{\em $^{2}$Department of Physics, University of Tennessee,
Knoxville, TN 37996--1200}} \\
{\small{\em $^{3}$ Oak Ridge National Laboratory, Oak Ridge, TN 37831--6373
 }} \\
\ \\
\ \\
\today
\end{center}
\vspace{3.7cm}

\begin{center}
{\large{\bf Abstract}}
\end{center}

\noindent
The  global correlation in the observed variation with mass number of the
$E2$ and summed $M1$ transition strengths is examined for rare earth nuclei.
It is shown that a theory of correlated $S$ and $D$ fermion pairs with a simple
pairing plus quadrupole interaction leads naturally to this universality.
Thus a unified and quantitative description emerges for low-lying quadrupole
and dipole strengths.

\newpage
\vspace{0.5cm}

When seemingly unrelated physical quantities exhibit similar empirical
behavior (a universality), one suspects a common physical genesis. One such
universality was recently recognized in nuclear structure physics: The measured
electromagnetic transition strength $B(E2)$ between the ground and the first
$2^+$ states of even--even rare earth nuclei exhibit a variation with mass
number $A$ that is similar to the variation with $A$ of the summed orbital
$B(M1)$ strengths measured in the same nuclei \cite{m1e2}.  In this letter, we
shall explore the physical implications of this universality.

The excited states associated with these  $E2$ and summed $M1$ strengths (the
$2^{+}_{1}$ and the lowest $1^{+}$ states, respectively) have very different
collective character. The former are symmetry states,
in which protons and
neutrons move with phase coherence; examples are low-lying rotations and
vibrations.  The latter are mixed-symmetry states,
in which protons and neutrons
move out of phase; examples are the lowest $1^{+}$\cite{1state} and some
excited $2^{+}$\cite{2state} states. Thus, the physical $0^{+}_{1} \rightarrow
1^{+}_{1}$ and $0^{+}_{1} \rightarrow 2^{+}_{1}$ transition strengths measure
two distinct classes of excitation: $E2$ excitations between symmetry states
and
$M1$ excitations between symmetry and mixed-symmetry states
(Although we address the
summed B(M1) strength of $1^{+}$ states up to 4 MeV in our calculations, the
dominant transition is typically $B(M1;0^{+}_{1} \rightarrow 1^{+}_{1})$).
Thus, the observed correlation in these two strengths hints at a microscopic
correlation for two modes which in the simplest geometrical picture appear to
be rather different. Its appearance provides a new window to gain additional
insights into the low energy behavior of nuclear structure.

Nuclear deformation is known to arise from the interplay of the
long-range n--p
quadrupole (QQ) interaction and the short-ranged pairing interaction between
like nucleons, as was emphasized by Federman and Pittel \cite {federman}.
However, straightforward application of the shell model for the nuclei in
question is prohibitively difficult. Algebraic approaches offer approximate
solutions of the shell model, and have provided some important insights into
this problem. For example, the neutron--proton version of the Interacting Boson
Model (IBM-2)\cite{IBM} allows one to use the $F$-spin \cite {Fspin} quantum
number to classify these states: symmetric with maximal $F$-spin and
mixed-symmetry with smaller values of $F$-spin. However,
Rangacharyulu et. al \cite{m1e2} concluded that it is difficult for the IBM-2,
at least under the assumption of preserving the $F$-spin symmetry, to account
for the universality. Likewise, Ginocchio \cite{gino} suggests that with the
F-spin symmetry and using the arguments from the well-known
Otsuka-Arima-Iachello (OAI) mapping procedure \cite {OAI} that it is not
straightforward to show both the $E2$ and $M1$ saturate well before midshell
(see the following figures for the data). Finally, $M1$ excitations have also
been specifically and quantitatively discussed with the quasiparticle picture
in the Nilsson model (e.g.  \cite{heyde}).

The Fermion Dynamical Symmetry Model (FDSM) \cite{fdsm,SO8} is an $SO(8)$ or
$Sp(6)$ truncation of the shell model.  Since Pauli correlations generally are
expected to become more important as midshell is approached, it is reasonable
to ask whether or not a model such as the FDSM that incorporates the true
fermionic nature of the correlated pairs can account for the similar behavior
of $E2$ and $M1$ strengths. We are encouraged to pursue this question by the
results from a recent paper \cite{exotic2} which showed that the same
$Q_{\pi}\cdot Q_{\nu}$ interaction is responsible for the splitting of symmetry
and the $2^{+}$ mixed-symmetry states,
and can consistently describe their transitional
properties. In this letter, we shall show that such a universality between $E2$
and $M1$ strengths arises within the FDSM framework with a reasonable set of
effective interaction parameters for the rare-earth nuclei.  In addition, we
shall demonstrate that the same calculation reproduces quantitatively the
energy ratio $E(4^{+}_{1})/E(2^{+}_{1})$, and accounts for the more subtle
deviations from universal behavior exhibited by these three quantities.

The primary building blocks of the FDSM are $S$ and $D$ (monopole and
quadrupole) correlated fermion pairs, whose integrity in low-lying states
is maintained even in the presence of the empirical single-particle energy
splittings \cite{SU3split}. The total neutron--proton FDSM symmetry for the
rare-earth nuclei is $Sp^{\nu}(6)\times SO^{\pi}(8)$. Ref.\ \cite{feng}
demonstrates that the low-lying spectroscopy of the rare earth nuclei
can be reasonably well described by a 5-parameter (pairing and QQ) FDSM
Hamiltonian:
\begin{equation}
 H=G'_{0\pi}S^{\dag}_{\pi}S_{\pi}^{\phantom{\dag}}+
    G'_{0\nu}S^{\dag}_{\nu}S_{\nu}^{\phantom{\dag}}+
    B'_{2\pi} P^{2}_{\pi} \cdot P^{2}_{\pi}+
    B'_{2\nu} P^{2}_{\nu} \cdot P^{2}_{\nu}+
    B_{2\pi\nu} P^{2}_{\pi} \cdot P^{2}_{\nu} .
\end{equation}
All the effective operators in Eq.(1) are defined in \cite{fdsm}. In this
Hamiltonian, the FDSM quadrupole pairing interactions are also taken into
account by renormalizing the parameters:
$G'_{0\sigma}=G_{0\sigma}-G_{2\sigma}$ and
$B'_{r\sigma}=B_{r\sigma}-G_{2\sigma}$ ($\sigma=\pi,\nu$).
One generally finds that the n--p QQ interaction is significantly stronger than
the pairing interaction between like nucleons.

The model space is restricted to the $S$--$D$ subspace in the normal-parity
shells (heritage $u=0$, corresponding to no broken pairs). Although the
particles in abnormal-parity levels are not included explicitly, they are
included effectively by the constraint that there is a distribution of
particles
between the normal and the
abnormal parity levels.
The number of pairs $(N_1$) in the
normal-parity levels is treated as a good quantum number and is calculated from
a semi-empirical formula determined globally from the
ground state spin of the odd-mass nuclei \cite{fdsm}.

In the IBM-2, the $F$-spin algebra allows the introduction of an $F$-spin
(Majorana) scalar interaction that has no effect on the
symmetric states.  Its strength is chosen to place
the mixed-symmetry states at the proper energies. The energies of the
low-lying symmetric $2^{+}_{1}$
and $4^{+}_{1}$ states are
usually chosen to determine the strength of the
$Q_{\pi}\cdot Q_{\nu}$ interaction. Thus in the
IBM-2, the strengths of the
$Q_{\pi}\cdot Q_{\nu}$ and the Majorana interactions are separately fitted to
states that have different collective behaviors. On the other hand,
the FDSM, and indeed most fermion models built from pairs,
cannot have a closed $F$-spin algebra.  This constraint will
prevent the fermion picture to having an analogus
phenomenological separation of and mixed-symmetry states.
Therefore, in the fermion picture,
one must describe all states, symmetric
or mixed-symmetric, by a single Hamiltonian.

The five parameters in Eq.\ (1) are determined numerically using a gradient
search within  the FDU0 code \cite{fdu0} to
best reproduce the experimental spectrum of the nuclide in question. The
experimental energies used in the fit for the systematics are $2^{+}_{1}$,
$4^{+}_{1}$, $6^{+}_{1}$, $1^{+}$, $2^{+}_{2}$, $0^{+}_{2}$ in Nd, Sm, Gd, Dy,
and Er (some $1^{+}$ states are not known experimentally  for Sm and Nd
isotopes).  This procedure is carried out until a good match to the
experimental spectra and a smooth trend in particle number was
found for the Hamiltonian parameters. Similar calculations
have been described in \cite{gui93g,fdsm} and details of the present
fitting process will be discussed in a forthcoming paper \cite{isop}.

When a suitable fit is found for a particular nuclide, the correlation of the
symmetry and mixed-symmetry states is given by the unified Hamiltonian of Eq.\
(1).
We find that the QQ-interaction plays the crucial role in
correlating the $2^{+}_{1}$ and $1^{+}$ states because
the excitation energies of these states depend sensitively on this term.
Once a suitable spectrum has been determined,
the next step is to see whether there
is correlation between symmetry and mixed-symmetry states caused by the same
$Q_{\pi}
\cdot Q_{\nu}$ strength for the electromagnetic transitions.

To compute the electromagnetic transitions, one needs the wavefunctions of the
states in question and the effective transitional operators. From our unified
fit, we are able to obtain these wavefunctions. The $M1$ and $E2$ effective
transition operators in the FDSM are
\begin{eqnarray}
T(M1)^{1}_{\mu} = \sqrt{\frac{3}{4\pi}}~(g_{\pi}L_{\pi} +g_{\nu}L_{\nu}), ~~~~
T(E2)^{2}_{\mu} = e_{\pi}P^{2}_{\mu}(i)_{\pi}+e_{\nu}P^{2}_{\mu}(k)_{\nu},
\end{eqnarray}
respectively, where
\begin{eqnarray}
& & P^{r}_{\mu}(i)=
\sqrt{5} \left[ b^{\dagger}_{ki}
\tilde{b} ^{\phantom{\dagger}}_{ki} \right] ^{0r}_{0\mu},~~
P^{r}_{\mu}(k)=
\sqrt{15/2} \left[ b^{\dagger}_{ki}
\tilde{b} ^{\phantom{\dagger}}_{ki} \right] ^{r0}_{\mu0},~~
r =1, 2, ~~~~   \nonumber   \\
& & L_{\pi}=\sqrt{5}P^{1}_{\sigma}(i)~~
L_{\nu}=\sqrt{8/3}P^{1}_{\sigma}(k)~  .
\end{eqnarray}
In the above, $e_{\pi}$ ($e_{\nu}$) is the proton (neutron) effective charge
and is fixed at 0.24 eb (0.20 eb)  globally for all the rare-earth nuclei
examined here. The g's
are the g-factors in the $S\mbox{-}D~$ subspace;
we take $g_{\pi}$=1.0 $\mu_{N}$ and $g_{\nu}$=0 for all nuclei examined here.

In Fig.\ 1, we plot the experimental and calculated energies of the $2^{+}_{1}$
and $1^{+}_{1}$ states as a function of the factor $P\equiv
N_{p}N_{n}/(N_{p}+N_{n})$ where $N_{p}$ ($N_{n}$) are the valence proton
(neutron) numbers, respectively. The $P$ scheme was introduced by
Casten et.\ al \cite{p,m1e2}) and can effectively display the global
systematics. In a separate paper \cite{isop}, we shall
discuss the detailed level structures of the rare-earth nuclei.
There we will show that nuclei with the same $P$ factor exhibit a correlation
between the ground-band structure and that of mixed-symmetry states
because they have the same $Q_{\pi}\cdot Q_{\nu}$ contribution, while prominent
discrepancies in the $\beta$ and $\gamma$ excitation states may be largely
attributed to the different pairing for protons and neutrons.

One can see the quality of the fit to the energies in Fig.\ 1. Notice that both
the $2^{+}_{1}$ and $1^{+}_{1}$ states show orderly variation with a nearly
constant energy gap between the two states (except for $^{154}$Gd). These
trends are also present in other states expected to be in the symmetric and
mixed-symmetric classes \cite{isop}.

The corresponding wavefunctions are used to compute the $E2$ and $M1$
transitions.The $B(E2)$ values and the summed $B(M1)$ strengths from this
calculation are shown in Figs.\ 2a' and 2b', while the corresponding data are
shown in Figs.\ 2a and 2b. The curves are the empirical relations presented in
Ref.\ \cite{m1e2} that summarize the approximate behavior of the data. The
$B(E2)$ and $B(M1)$ strengths are reproduced quantitatively by the
calculations. Thus, we find theoretical evidence for the approximate  universal
behavior of $E2$ and $M1$ strengths exhibited by the data.  Furthermore, we
observe that even the deviations from universality exhibited by the data (for
example, the $M1$ saturation is sharper and occurs at least 1 unit of $P$ lower
than that for the $E2$ strength (which never completely saturates)), is
reproduced {\em quantitatively} by the calculations without parameter
adjustment.

In Figs. 2c and Figs. 2c',  we have plotted the ratio
$E(4^{+}_{1})/E(2^{+}_{1})$ as a function of $P$.  This quantity is also seen
to exhibit an empirical variation with $P$ that is similar to that of the
$E2$ and $M1$ strengths, and it is also quantitatively reproduced by these
calculations. Since we expect this ratio to be sensitive to the $Q_{\pi} \cdot
Q_{\nu}$ interaction, this is an expected result given the success of the
preceding calculations and our previous assertion that the
$Q_{\pi} \cdot Q_{\nu}$ term is the most important factor governing the
relationship between the properties of the symmetry and mixed-symmetry states.

Finally, we address the question of how important the variation of effective
interaction parameters is to the success of these calculations.
In Figs.\ 2a''--2c'' we repeat the calculations of Figs.\ 2a'--2c', but with a
{\em fixed set of parameters for all nuclei:}
$G'_{0\pi}$=-0.074 MeV, $G'_{0\nu}$=0.020 MeV, $B'_{2\pi}$=-0.001 MeV,
$B'_{2\nu}$=0.047 MeV, and $B_{2\pi\nu}$=-0.243 MeV. Remarkably, these
calculations are also in quantitative agreement with observations, differing
only in minor details from the previous calculations in which the effective
interactions have a weak $A$ dependence. Thus, the quantitative reproduction of
$E2$ and $M1$ strengths in the rare  earth nuclei is an inherent feature of the
FDSM: we reiterate that in Fig.\ 2, no parameters have been adjusted to either
the $E2$ or $M1$ strengths. It should be noted that the small positive value of
the renormalised $G'_{0\nu}$ means that the neutron quadrupole pairing is
strong, thus implying that higher angular momentum pairs may also play some
role.

Let us note that the successes of these calculations depend on the
separation of particles into abnormal and normal-parity orbitals in the FDSM
valence space, with only the particles in the normal-parity orbitals
contributing directly to the $Sp(6) \times SO(8)$ collectivity.
This separation has been specified by prior considerations
wholly unrelated to the present discussion of $E2$ and $M1$ strengths.
That the resulting theory describes quantitatively the global behavior of
these collective strengths without parameter adjustment is evidence
of the FDSM assumption that normal parity and abnormal
parity orbitals play fundamentally different roles.
Evidence for this separation has been presented before in the systematics
of nuclear masses and deformation \cite{han92,gui93f,fdsm}.
Such a division of the roles of normal and abnormal
parity orbitals is not within the framework of the IBM, but its effect
would be to lower the effective $d$ boson number in the region of deformed
nuclei. Perhaps adjusting the $d$-boson number may be an empirical way to
obtain an IBM-2 description of these systematics.

In summary, the observed approximate universal behavior of nuclear collective
symmetry and mixed-symmetry states is examined here. The $F$-spin formalism of
the
Interacting Boson Model suggests an elegant classification of
these states, but an IBM-2 model with $F$-spin symmetry may not easily
account for the universality exhibited by the corresponding $M1$
and $E2$ strengths. It is thus interesting to inquire
whether by allowing $F$-spin mixing the IBM-2 can address the
correlation between the $M1$ and $E2$ strengths.
What we have demonstrated here is that the
FDSM can quantitatively reproduce the $M1$ and $E2$ strengths
as well as the energy
ratio $E(4^{+}_{1})/E(2^{+}_{1})$: the calculations can reproduce
both the approximate universality and its subtle deviations
without parameters adjusted to either the $M1$ or $E2$ strengths.
This suggests that a properly chosen n--p quadrupole interaction in a
symmetry-truncated
fermion model can simultaneously account for the properties of
symmetry and mixed-symmetry states. These results depend non-trivially on the
separation of normal and abnormal parity orbitals, and represent another
piece of evidence for this separation.

We believe the present conclusions to be rather general in nature.
The FDSM illustrates the physics transparently and economically, but
{\em any} fermion model leading to a good
collective subspace and with realistic effective interactions suited to that
subspace should be able to account for this behavior of the collective
strengths. It will be important to see whether other fermion models can
accommodate the approximate universality of
symmetry and mixed-symmetry states in as
simple a manner as presented here.

This research is supported by the NSF (Drexel) and DOE (UT and ORNL).

\newpage
{\bf \large Figure Caption}

\noindent
Fig.\ 1  \vspace{5pt}
Experimental and calculated energy levels
for the first $2^+$ (lower points) and $1^+$ (upper
points) states in selected even--even rare earth nuclei where orbital M1
strengths have been measured.  The symbols denote isotopes with the same
meanings as in Fig.\ 2.

\vspace{0.1cm}
\noindent
Fig.\ 2  \vspace{5pt} Comparison of experimental and theoretical
strenghs for $B(E2)$ and summed $B(M1)$ for the rare-earths. Both are
plotted as functions of $P$. The data which appear in
the left column are taken from
\cite{m1e2} ,\cite{raman} and \cite{new}.
The points with different symbols
in (a')-(c') and
(a")-(c") are theoretical results for different isotopes. For comparison,
curves from the same empirical relation are used here, namely
$B(E2,M1)=a_{1}+a_{2}/[1+exp((c-P)/d)]$ (In (a)--(a''),
$a_{1}$=1.3, $a_{2}$=1.1,
$c$=5.45 and $d$=0.57. In (b)--(b''), $a_{1}$=0.36, $a_{2}$=2.2.,
$c$=4.1 and $d$=0.32. In (c)--(c''), $a_{1}$=1.3, $a_{2}$=1.9.,
$c$=3.3 and $d$=0.49. ) are also plotted. The symbols in (c)--(c'')
have the same meaning as in the $E2$ and $M1$ cases.
The theoretical results in column 2 correspond to the parameters
which can best fit the data. Finally,
the theoretical results of column
3 correspond to constant values of the effective interaction parameters.

\end{document}